\documentclass[amssymb,11pt]{article}
\usepackage{amsmath}
\usepackage{amssymb}
\usepackage{graphicx}
\newtheorem{thm}{Theorem}[section]

\newtheorem{lem}{Lemma}[section]
\newtheorem{defn}{Definition}[section]
\newtheorem{prop}{Proposition}[section]

\begin{document}
\title{Another Quantum Lov\'{a}sz Local Lemma}
\author{Mingsheng Ying\thanks{
This work was partly supported by the Australian Research Council
(Grant No: DP110103473) and the National Natural Science Foundation
of China (Grant No: 60736011).}\\
\\
\small \em Center for Quantum Computation and Intelligent Systems,\\
\small \em Faculty of Engineering and Information Technology,\\
\small \em University of Technology, Sydney,\\ \small \em NSW 2007,
Australia\\ \small \em and\\
\small \em State Key Laboratory of Intelligent Technology and Systems,\\
\small \em Tsinghua National Laboratory for Information Science and
Technology,\\
\small \em Department of Computer Science and Technology,\\
\small \em Tsinghua University, Beijing 100084, China,\\
\small \em Email: mying@it.uts.edu.au, yingmsh@tsinghua.edu.cn}
\date{}
\maketitle

\begin{abstract}

We define a natural conceptual framework in which a generalization
of the Lov\'{a}sz Local Lemma can be established in quantum
probability theory.

\smallskip\

\end{abstract}\par

\vspace{1em}

\section{INTRODUCTION}

The idea of the probabilistic method~\cite{AS04} can be roughly
described as follows: in order to prove the existence of an object
with a certain desired property, one finds an appropriate
probability space of objects and then shows that the required
property holds with a positive probability. It is quite often that
the object under consideration is \textquotedblleft rare" in the sense that its
desired property is the combination of a large number of basic
properties. If these basic properties are independent of one
another, one can prove the existence of the object by verifying the
possibility of each basic property; but it is usually difficult to
handle a combinations of multiple properties without the
independence condition. Fortunately, the Lov\'{a}sz Local Lemma (LLL
for short)~\cite{EL75} provides a powerful tool for dealing with the
case of \textquotedblleft limited" dependence. Formally, it can be stated as the
following:

\begin{thm} (Symmetric Lov\'{a}sz Local Lemma~\cite{EL75})

Let $E_1,...,E_n$ be events over some sample space such that
$Pr(E_i)\leq p$ for all $1\leq i\leq n$, and each of them is
mutually independent of all but $d$ of the others. If $p\cdot e\cdot
(d+1)\leq 1$, then $Pr(\bigwedge_{i=1}^{n}\overline{E_i})>0$.
\end{thm}

The LLL has found a lot of applications in graph theory,
combinatorics, algorithms and complexity, etc., and a remarkable one
is to prove the existence of solutions to the $k-$SAT problem.
Recently, motivated by the local Hamiltonian problem~\cite{KKR04} in
quantum mechanics, Bravyi~\cite{Br06, BMR10} defined a quantum
analogue of $k-$SAT, called $k-$QSAT. To offer a tool similar to the
LLL for $k-$QSAT, Ambainis, Kempe and Sattath~\cite{AKS10}
discovered a quantum generalization of the LLL. Their quantum LLL is
obtained by a classical-quantum correspondence of the basic notions
in probability theory, e.g. probability space, events, probability,
conditional probability and independence. A (finite-dimensional)
Hilbert space $\mathcal{H}$ is corresponding to a probability space,
and its subspaces are seen as events. The probability of event $X$
is defined to be its relative dimension $R(X)=\frac{\dim X}{\dim
\mathcal{H}}.$ Then the notions of conditional probability and
(R-)independence can be introduced in a familiar way, and the following
is derived:

\begin{thm}\label{AKS} (Symmetric Relative Dimension Quantum Lov\'{a}sz Local Lemma~\cite{AKS10})

Let $X_1,...,X_n$ be subspaces of a finite-dimensional Hilbert space
such that $R(X_i)\geq 1-p$ for all $1\leq i\leq n$, and each
subspace is mutually $R-$independent of all but $d$ of the others.
If $p\cdot e\cdot (d+1)\leq 1$, then $R(\bigcap_{i=1}^{n}X_i)>0$.
\end{thm}

The above quantum LLL was immediately proved to be very powerful by
using it to considerably improve some results in $k-$QSAT and random
QSAT~\cite{LLMSS09, LMSS09}. In particular, it can be used to assert
the existence of entangled state solutions to the QSAT problems,
whereas previous approaches can only find product states. As already
pointed out in~\cite{AKS10}, Theorem~\ref{AKS} is called a quantum
LLL mainly because of its successful applications in some quantum
problems. There is nothing quantum even probabilistic in itself.
Indeed, it is a statement about the relative dimensions of subspaces
of a Hilbert space, and can be more appropriately called a
\textquotedblleft geometric" LLL.

This paper follows the research line of seeking a quantum version of
the LLL initiated in~\cite{AKS10}. We aim at finding a
generalization of the LLL in a general framework of quantum
probability. To this end, a key step is still to look for a
classical-quantum correspondence of the notions involved in the LLL.
Indeed, such a correspondence has already been well-established in
the theory of quantum measurements: A density operator defines a
probability space. An event can be seen as a constraint that the outcome of a measurement
lies in a prescribed range. The probability of an event is then given
according to the basic postulate of quantum mechanics for
measurements. As early as in 1965, Hautappel, van Dam and Wigner~\cite{HDW65}
introduced the so-called $\Pi$ function in their treatment of geometric invariance principles. The $\Pi$
function gives exactly the conditional probability of a sequence of events after the realization of another sequence of events
when all the involved measurements are projective. The notion of conditional probability was later introduced by Davis~\cite{Da76} in the case of general quantum measurements.  
Furthermore, the notion of conditional probability can be employed to define independence between quantum events in a familiar way. 

The key thing that we have to carefully consider is the ordering of multiple events when they are combined by means of \textquotedblleft and". It does not matter in the classical probability theory because the connective \textquotedblleft and" is commutative in Boolean logic. Nevertheless, the ordering of quantum events plays a decisive role due to non-commutativity of measurements, which is a basic
fact in quantum mechanics. Keeping this in mind, a quantum LLL can be then established based on a general notion of
quantum measurements by respecting the ordering of events. The remaining difficulty comes from the monotonicity of probability:\begin{equation}\label{mono}
Pr[\bigwedge_{i=1}^{n}E_i]\leq Pr[E_1\wedge...\wedge E_{k-1}\wedge E_{k+1}\wedge ...\wedge E_n],
\end{equation} which is required in the proof of LLL. In the classical case, the monotonicity can be simply understood that loosening a constrain $\bigwedge_{i=1}^{n}E_i$ by deleting $E_k$ increases its probability. In the quantum case, however, there are two ways to remove the event $E_k$: (a) we do not perform the measurement that defines event $E_k$; and (b) we do perform the measurement but put no constraint on its outcome (i.e. the outcome is allowed to be anything). Then the remaining difficulty can be easily resolved by understanding the subtle difference between (a) and (b): the quantum counterpart of Eq.~(\ref{mono}) is not true unless $k=n$ when $E_k$ is removed according to (a), but it is always valid if (b) is adopted.  

This paper is organized as follows. The notion of a test is introduced in Sec.~\ref{condi} in order to define quantum probability with monotonicity. Sec.~\ref{condi}  is devoted to carefully
examine basic properties of conditional probability in a test,
compared with conditional probability in a state. The notion of
independence in a test is also introduced in Sec.~\ref{condi}. Quantum generalizations of both the general (asymmetric) LLL and the symmetric LLL are proved in Sec.~\ref{ql}. The LLL and its quantum generalization appear in the same way except a minor difference: for the LLL, \textquotedblleft limited" dependence means that each event only (probabilistically) depends on a small number of other events, whereas for the quantum LLL, it means that each event only depends on the events which are not far from it in the sequence of all the events under consideration.

\section{QUANTUM CONDITIONAL PROBABILITY}\label{condi}

Recall from~\cite{NC00} that a quantum measurement in a Hilbert
space $\mathcal{H}$ is a family $M=\{M_m\}$ of linear operators on
$\mathcal{H}$ such that $\sum_m M_m^{\dag}M_m=I_\mathcal{H},$ where
$I_\mathcal{H}$ is the identity operator on $\mathcal{H}$. The index
$m$ stands for the outcome of measurement, and it ranges over a set
$spec(M)$, called the spectrum of $M$.

\begin{defn} An event in a Hilbert space $\mathcal{H}$ is a pair
consisting of a quantum measurement $M$ in $\mathcal{H}$ and a
subset $A$ of the spectrum $spec(M)$ of $M$. We will write $E=\{M\in
A\}$ for such an event. The intuitive meaning of this event is that
the outcome of measurement $M$ lies in $A$. We often say that $E$ is defined by $M$.\end{defn}

For simplicity, we will write $\{M=m\}$ for event $\{M\in A\}$
whenever $A$ is a singleton $\{m\}$. If a positive operator $\rho$ satisfies $tr(\rho)\leq 1$, then $\rho$ is called a partial density operator.
For any event $E=\{M\in A\}$,
we define a super-operator $\mathcal{E}$ as follows: for any partial
density operators $\rho$,
\begin{equation}\label{def-sup}\mathcal{E}(\rho)=\sum_{m\in A}M_m\rho M_m^{\dag}.\end{equation} Obviously,
$\mathcal{E}(\rho)$ is also a partial density operator, and
super-operator $\mathcal{E}(\cdot)$ enjoys linearity:
$\mathcal{E}(\sum_ip_i\rho_i)=\sum_ip_i\mathcal{E}(\rho_i)$ for all density operators $\rho_i$ and real numbers $p_i\geq 0$ with $\sum_ip_i\leq 1$. 

\begin{defn}\begin{enumerate}\item Let $M$ be a measurement. Then
events $\emptyset_M=\{M\in\emptyset\}$ and $I_M=\{M\in spec(M)\}$
are called the empty and complete events, respectively, defined by $M$
\item
For any event $E=\{M\in A\}$, its complement is defined to be
$\overline{E}=\{M\in spec(M)\setminus A\}$. Intuitively,
$\overline{E}$ means that the outcome of measurement does not lie in
$A$.

\item For
any two events $E_1=\{M\in A_1\}$ and $E_2=\{M\in A_2\}$ defined by the
same measurement, their union is defined to be $E_1\cup E_2=\{M\in
A_1\cup A_2\}$. Intuitively, $E_1\cup E_2$ means that the outcome of
measurement lies either in $A_1$ or in $A_2$.
\end{enumerate}
\end{defn}

It is easy to see that the super-operator defined by empty event
$\emptyset_M$ according to Eq.~(\ref{def-sup}) is the null operator:
$\mathcal{E}(\rho)=0$ for all $\rho$. We write $\mathcal{I}_M$ for
the super-operator defined by complete event $I_M$. If $E$ is
defined by measurement $M$, and the super-operator defined by $E$
and $\overline{E}$ are $\mathcal{E}$, $\overline{\mathcal{E}}$,
respectively, then
$\mathcal{E}+\overline{\mathcal{E}}=\mathcal{I}_M$. Let $E_i=\{M\in
A_i\}$ $(i=1,2)$, and let $\mathcal{E}_1$, $\mathcal{E}_2$ and
$\mathcal{E}$ be the super-operators defined by $E_1$, $E_2$ and
$E_1\cup E_2$, respectively. If $A_1\cap A_2=\emptyset,$ then
$\mathcal{E}_1+\mathcal{E}_2=\mathcal{E}$. It should be noted that
in general $\mathcal{I}_M$ is not the identity operator. Moreover,
it is possible that $\mathcal{I}_{M_1}\neq\mathcal{I}_{M_2}$ when $M_1$
and $M_2$ are two different measurements; for example, let
\begin{equation}\label{ex-mea}\begin{split}M_1&=\{M_{10}=|0\rangle\langle 0|,M_{11}=|1\rangle\langle
1|\},\\ M_2&=\{M_{20}=|+\rangle\langle +|,M_{21}=|-\rangle\langle
-|\},\end{split}\end{equation} 
where $|\pm\rangle=\frac{1}{\sqrt{2}}(|0\rangle+|1\rangle)$. 
If $\rho=|0\rangle\langle 0|$, then
$\mathcal{I}_{M_1}(\rho)=|0\rangle\langle 0|\neq
\frac{1}{2}(|+\rangle\langle +|+|-\rangle\langle
-|)=\mathcal{I}_{M_2}(\rho)$.

\subsection{Quantum Probability in a State}\label{state}

Suppose that a physical system is in state $\rho$, where $\rho$ is a
density operator in Hilbert space $\mathcal{H}$. Then the
probability of event $E=\{M\in A\}$ in state $\rho$ is
\begin{equation}\label{pro-rho}Pr_\rho[E]=\sum_{m\in A}tr(M_m\rho
M_m^{\dag})\end{equation} according a basic principle of quantum
mechanics. Equivalently, $Pr_\rho(E)=tr(\mathcal{E}(\rho))$, where
$\mathcal{E}$ is the super-operator defined by event $E$ according
to Eq.~(\ref{def-sup}). 
More generally, let $\mathbb{E}=E_1,E_2,...,E_k$ be a sequence of
events. We will write $|\mathbb{E}|$ for the length of $\mathbb{E}$, i.e. $|\mathbb{E}|=k$. For each $1\leq i\leq k$, we write $\mathcal{E}_i$ for the super-operator defined by $E_i$ according to Eq.~(\ref{def-sup}), and let $\mathcal{E}$ be the composition of $\mathcal{E}_1, \mathcal{E}_2,...,\mathcal{E}_k$, i.e. $\mathcal{E}=\mathcal{E}_k\circ ...\circ \mathcal{E}_2\circ\mathcal{E}_1$. Then the probability of $\mathbb{E}$ in state $\rho$ is given by
\begin{equation}\label{pr-seq}Pr_\rho[\mathbb{E}]
=tr((\mathcal{E}(\rho))=tr[\mathcal{E}_k(...\mathcal{E}_2(\mathcal{E}_1(\rho))...)].\end{equation}Intuitively, if $E_i=\{M_i\in A_i\}$ for all $1\leq i\leq n$, then
Eq.~(\ref{pr-seq}) means that measurement $M_1$ is first performed
and the outcome lies in $A_1$, and then $M_2$ is performed and the
outcome lies in $A_2$, and so on, until $M_k$ is performed and the
outcome lies in $A_k$. It is worth noting that the ordering of
performing these measurements is fixed. 

For any two sequences $\mathbb{E}$ and $\mathbb{F}$ of events, we write $\mathbb{E},\mathbb{F}$ for the concatenation of $\mathbb{E}$ and $\mathbb{F}$, i.e. if $\mathbb{E}=E_1,...,E_k$ and $\mathbb{F}=F_1,...,F_l$, then $\mathbb{E},\mathbb{F}=E_1,...,E_k,F_1,...,F_l$. Here, either $\mathbb{E}$ or $\mathbb{F}$ is allowed to be a single event.  

\begin{prop}\label{property}\begin{enumerate}\item If all the events in $\mathbb{E}$ are defined by the same projective measurement, and $\mathbb{E}^{\prime}$ is a permutation of $\mathbb{E}$, then $Pr_\rho[\mathbb{E}]=Pr_\rho[\mathbb{E}^{\prime}]$.  
\item $Pr_\rho[\mathbb{E}, \emptyset_M,\mathbb{F}]=0$ for any $\mathbb{E}$ and $\mathbb{F}$. \item If $\mathbb{I}$ is a sequence of complete events that can be defined by different measurements, then $Pr_\rho[\mathbb{E},\mathbb{I}]=Pr_\rho[\mathbb{E}]$; in particular, $Pr_\rho[\mathbb{I}]=1$.  \item $Pr_\rho[\mathbb{E},\overline{E}]=Pr_\rho[\mathbb{E}]-Pr_\rho[\mathbb{E},E]$.\item If $E_i=\{M\in A_i\}$ $(i=1,2)$ with $A_1\cap A_2=\emptyset$, then $Pr_\rho[\mathbb{E},E_1\cup E_2,\mathbb{F}]=Pr_\rho[\mathbb{E},E_1,\mathbb{F}]+Pr_\rho[\mathbb{E},E_2,\mathbb{F}]$.
\end{enumerate}
\end{prop}

The equality $Pr_\rho[\mathbb{E}]=Pr_\rho[\mathbb{E}^{\prime}]$ in clause 1 of the above proposition is not valid in general; for example, let $M_1$ and $M_2$ be given by Eq.~(\ref{ex-mea}) and $\rho=|+\rangle\langle +|$. Then $Pr_\rho[\{M_1=1\},\{M_2=0\}]=\frac{1}{4}\neq 0=Pr_\rho[\{M_2=0\},\{M_1=1\}]$. Clause 3 indicates that complete events occurring in the last part of a sequence of events can be removed. However, complete events in the other part cannot be removed, as shown by the following simple example: $Pr_\rho[I_{M_1},\{M_2=0\}]=\frac{1}{2}<1=Pr_\rho[\{M_2=0\}]$ and $Pr_\rho[I_{M_1},\{M_2=1\}]=\frac{1}{2}>0=Pr_\rho[\{M_2=1\}]$. This example also shows that clause 4 is not true whenever $E$ and $\overline{E}$ do not occur after $\mathbb{E}$ but are inserted in the middle of $\mathbb{E}$, since $Pr_\rho[I_{M_1},\{M_2=0\}]=Pr_\rho[E,\{M_2=0\}]+Pr_\rho[\overline{E},\{M_2=0\}]$, where $E=\{M_1=0\}$.

The notion of conditional probability can also be defined in a
familiar way. Given a density operator $\rho$ in Hilbert space
$\mathcal{H}$. The conditional probability of a sequence
$\mathbb{F}$ of events after a sequence $\mathbb{E}$ of
events in state $\rho$ is defined to be
\begin{equation}\label{def-cond}Pr_\rho[\mathbb{F}|\mathbb{E}]
=\frac{Pr_\rho[\mathbb{E},\mathbb{F}]}{Pr_\rho[\mathbb{E}]}.\\
\end{equation} If we only consider projective measurements, then the
conditional probability defined in Eq.~(\ref{def-cond}) is exactly
the Houtappel-van Dam-Wiger's $\Pi-$function in the case of discrete
time~\cite{HDW65}. For the case of general measurements, it was
introduced by Davis~\cite{Da76}.

Some basic properties of quantum conditional probability in a state are
collected in the following:

\begin{prop}\label{con-p}\begin{enumerate}\item If $M$ is a projective measurement, then for any $A$, we have
$Pr_\rho[\{M\in A\}|\{M\in A\}]=1.$
\item Monotonicity:
$Pr_\rho[\mathbb{F},\mathbb{G}|\mathbb{E}]\leq
Pr_\rho[\mathbb{F}|\mathbb{E}].$
\item Additivity: If $E_i=\{M \in A_i\}$ $(i=1,2)$ with $A_{1}\cap A_{2}=\emptyset$, then
$$Pr_\rho[\mathbb{F},E_1\cup E_2,\mathbb{G}|\mathbb{E}]= Pr_\rho[\mathbb{F},E_1,\mathbb{G}|\mathbb{E}]+Pr_\rho[\mathbb{F},E_2,\mathbb{G}|\mathbb{E}].$$ In particular,
$Pr_\rho[F|\mathbb{E}]
+Pr_\rho[\overline{F}|\mathbb{E}]=1.$
\item Chain Rule:
\begin{equation*}Pr_\rho[F_1,...,E_l|\mathbb{E}]
=\prod_{i=1}^{l}Pr_\rho[F_i|\mathbb{E}, F_1,...,F_{i-1}].
\end{equation*}\end{enumerate}
\end{prop}

The monotonicity given in Proposition~\ref{con-p}.2 indicates that
deleting the tail $\mathbb{G}$ of the sequence $\mathbb{F},\mathbb{G}$ in conditional probability
$Pr_\rho [\mathbb{F},\mathbb{G}|\mathbb{E}]$ does not
decrease the value of the probability. It is well-known that in
classical probability theory the same happens when we delete some events in the other part. However, it is not the case in quantum probability
theory, as shown by the following example: let
$\rho=|+\rangle\langle +|$, $M_1$ and $M_2$ be given by
Eq.~(\ref{ex-mea}), $M_3=\{M_{30}=|0\rangle\langle 0|,
M_{31}=|1\rangle\langle 1|\},$ $E_1=\{M_1=0\}$, $E_2=\{M_2=0\}$ and
$E_3=\{M_3=1\}$. Then $Pr_\rho
[E_2,E_3|E_1]=\frac{1}{4}>0=Pr_\rho[E_3|E_1].$ Unfortunately, the
monotonicity of conditional probability with respect to the head $\mathbb{F}$ of the sequence $\mathbb{F},\mathbb{G}$ is essential in the proof of LLL.
So, the conditional probability defined in Eq.~(\ref{def-cond})
cannot directly be used to establish a quantum generalization of
LLL. The same example shows that the following total probability law
does not hold:
\begin{equation*}Pr_\rho[\mathbb{E},\mathbb{F},\mathbb{G}]
=\sum_i Pr_\rho[\mathbb{E},\{M\in A_i\},\mathbb{F}]\cdot
Pr_\rho[\mathbb{G}|\mathbb{E},\{M\in
A_i\},\mathbb{F}],
\end{equation*}where $\{A_i\}$ is a partition of
$spec(M)$. In fact,
$Pr_\rho[E_1,E_3]=0<\frac{1}{4}=Pr_\rho[E_1,E_2,E_3]+Pr_\rho[E_1,\overline{E_2},E_3]=\sum_{i=0}^{1}
Pr_\rho[E_3|E_1,\{M_2=i\}]\cdot Pr_\rho[E_1,\{M_2=i\}].$

\subsection{Quantum Probability in a Test}

\begin{defn}A test in Hilbert space $\mathcal{H}$ is a tuple $\Sigma=(\rho;M_1,M_2,...,M_n)$,
where $\rho$ is a density operator in $\mathcal{H}$, $n\geq 1$, and
$M_1,M_2,...,M_n$ are measurements in $\mathcal{H}$. Intuitively,
$\Sigma$ means that we prepare a physical system in state $\rho$,
and then perform measurements $M_1$ through $M_n$ on this system.
\end{defn}

It should be pointed out that the ordering of the measurements in an
test $\Sigma=(\rho;M_1,M_2,...,M_n)$ is fixed: we first perform
$M_1$, then $M_2$, and so on; finally, we perform $M_n$. The
probabilities considered in Subsec. 2.1 are defined in a physical
system in a given state $\rho$. Now, we are going to define
probabilities in a given test, where not only the initial state of
the physical system but also a sequence of measurements to be
performed on the system are specified. First, we fix some notations. For any positive integer $k$, we write $(k]$ for the sequence $1,2,...,k$ of smallest positive integers up to $k$. If $K=i_1,i_2,...,i_k$ is a subsequence of $(n]$, then we use $\max K$ and $\min K$ to denote the last (and greatest) element $i_k$ and the first (and smallest) element $i_1$, respectively, of $K$. If $\max K<i$, then we write $K<i$. If $L$ is another subsequence of $(n]$ such that $\max K<\min L$, then we write $K<L$. In this case, we write $K,L$ for the concatenation of $K$ and $L$. For any subsequence $J$ of $K$, $K\setminus J$ stands for the complement of $J$ in $K$; that is, the resulting sequence of deleting all the elements of $J$ from $K$.    
Also, we will write $\mathbb{E}_K$ for the subsequence $E_{i_1},E_{i_2},...,E_{i_k}$ of sequence $\mathbb{E}=E_1,E_2,...,E_n$ of events.   

\begin{defn}Let $\Sigma=(\rho;M_1,M_2,...,M_n)$ be a test. \begin{enumerate}\item
If $k\leq n$, and $E_i$ is an event defined by measurement $M_i$ for
each $1\leq i\leq k$, then the (joint) probability of sequence
$\mathbb{E}_{(k]}$ of events in $\Sigma$ is defined by
\begin{equation}\label{def-prob}Pr_\Sigma [\mathbb{E}_{(k]}]=Pr_\rho [\mathbb{E}_{(k]}].\end{equation} Intuitively,
the sequence $\mathbb{E}_{(k]}$ can be understood as a sequential
composition of events $E_1,E_2,...,E_k$. More precisely, if
$E_i=\{M_i\in A_i\}$ for every $i\leq k$, then this sequence means
that measurement $M_1$ is first performed and the outcome lies in
$A_1$, and then $M_2$ is performed and then outcome lies in $A_2$,
and so on, until $M_k$ is performed and the outcome lies in $A_k$.
\item If $K$ is a subsequence of $(n]$, and $E_{i}$ is an event defined by measurement $M_{i}$ for each element $i$ of $K$, then the
(marginal) probability of sequence $\mathbb{E}_K$ of
events in $\Sigma$ is defined by
\begin{equation}\label{def-mar}Pr_\Sigma [\mathbb{E}_K]=
Pr_\Sigma
[\mathbb{E}^{\prime}_{(\max K]}],\end{equation}
where $E_i^{\prime}=I_{M_i}$ for $i\in (\max K]\setminus K$, and
$E_{i}^{\prime}=E_{i}$ for $i\in K.$ In particular,
for any $1\leq i\leq n$, we define $Pr_\Sigma
[E_i]=Pr_\Sigma[I_{M_1},...,I_{M_{i-1}},E_i].$
\end{enumerate}
\end{defn}

It is obvious that if $K$ is an initial segment $(k]$ of
the sequence $(n]$, 
then Eq.~(\ref{def-mar}) degenerates to Eq.~(\ref{def-prob}). However,
if $K$ is not an initial segment of $(n]$, then the defining equation
of $Pr_\Sigma [\mathbb{E}_K]$ deserves a careful
explanation. Since the ordering of measurements $M_1,M_2,...,M_n$ is
fixed in the test $\Sigma$, it is not allowed to perform measurement
$M_j$ if $M_i$ was not performed whenever $1\leq i<j\leq n$. This is
why we have to insert complete events $I_{M_i}$ in Eq.~(\ref{def-mar}) for
all $i\in (\max K]\setminus K\neq\emptyset$. 
So, Eq.~(\ref{def-mar}) indicates that we have to perform all of
measurements $M_1,M_2,...,M_{\max K-1},M_{\max K}$ but not only measurements
$M_{i}, i\in K$. It does not hold that
$Pr_\Sigma [\mathbb{E}_K]=Pr_\rho [\mathbb{E}_K]$ in general.
In particular, it may happen that $Pr_\Sigma[E_i]\neq Pr_\rho[E_i]$
for $i>1$. For example, consider test $\Sigma=(\rho;M_1,M_2)$, where
state $\rho=|+\rangle\langle +|$, and measurements $M_1$ and $M_2$
are given by Eq.~(\ref{ex-mea}). Let event $E_2=\{M_2=0\}$. Then
$Pr_\rho[E_2]=1>\frac{1}{2}=Pr_\Sigma[I_{M_1},E_2]=Pr_\Sigma[E_2].$
On the other hand, if $E_2^{\prime}=\{M_2=1\}$, then
$Pr_\rho[E_2^{\prime}]=0<\frac{1}{2}=Pr_\Sigma[E_2^{\prime}]$.

Based on the above definition, the notion of conditional probability
can be introduced in a natural way.

\begin{defn}Given a test $\Sigma=(\rho;M_1,M_2,...,M_n)$. Let $K$ and $L$ be two subsequences of $(n]$ with $K<L$, and let $E_{i}$ be
an event defined by measurement $M_{i}$ for each $i\in K,L$. Then the conditional probability of sequence
$\mathbb{E}_L$ of events after sequence
$\mathbb{E}_K$ of events in $\Sigma$ is defined to be
\begin{equation}\label{def-cond1}Pr_\Sigma [\mathbb{E}_L|\mathbb{E}_K]
=\frac{Pr_\Sigma [\mathbb{E}_K,\mathbb{E}_L]}{Pr_\Sigma [\mathbb{E}_K]}.\end{equation}\end{defn}

It should be noted that measurements $M_i$ for $1\leq i\leq
\max (K,L)$ but not only those $M_{i}$ with $i\in K,L$ are performed
in the above defining equation of conditional probability. In general, it does not hold that
$Pr_\Sigma [\mathbb{E}_L|\mathbb{E}_K]=Pr_\rho [\mathbb{E}_L|\mathbb{E}_K]
$ unless the concatenation $K,L$ of $K$ and $L$ is an initial segment of $(n]$.

Some basic properties of quantum conditional probability in a test are
presented in the following:

\begin{prop}\label{con-p1}Let $\Sigma=(\rho;M_1,M_2,...,M_n)$ be a
test, and let $E_1,E_2,...,E_n$ be a sequence of events such that
$E_i$ is defined by measurement $M_i$ for each $1\leq i\leq n$.
\begin{enumerate}\item If $1\leq i<n$, $M_i$ and $M_{i+1}$ are the same projective measurement $M$,
and $E_i=E_{i+1}=\{M\in A\}$ for some $A\subseteq spec(M)$, then
$Pr_\Sigma[E_{i+1}|E_i]=1.$
\item Monotonicity:  $Pr_\Sigma[\mathbb{E}_L|\mathbb{E}_K]\leq Pr_\Sigma[\mathbb{E}_J|\mathbb{E}_K]$ for any subsequence $J$ of $L$. 
\item Additivity: If $K,L,J$ are subsequences of $(n]$ and $1\leq i\leq n$ such that $J<K<i<L$, and $E_i^{(t)}=\{M_i\in A_t\}$ $(t=1,2)$ with $A_1\cap A_2=\emptyset$, then 
$$Pr_\Sigma[\mathbb{E}_K,E_i^{(1)}\cup E_i^{(2)},\mathbb{E}_L|\mathbb{E}_J]=
Pr_\Sigma[\mathbb{E}_K,E_i^{(1)},\mathbb{E}_L|\mathbb{E}_J]+Pr_\Sigma[\mathbb{E}_K, E_i^{(2)},\mathbb{E}_L|\mathbb{E}_J].
$$ In particular, we have:
$$Pr_\Sigma[\mathbb{E}_K,E_i,\mathbb{E}_L|\mathbb{E}_J]+
Pr_\Sigma[\mathbb{E}_K,\overline{E_i},\mathbb{E}_L|\mathbb{E}_J]=Pr_\Sigma[\mathbb{E}_K,\mathbb{E}_L|\mathbb{E}_J].
$$ Furthermore, if $J<j$, then $Pr_\Sigma[E_j|\mathbb{E}_J]+Pr_\Sigma[\overline{E_j}|\mathbb{E}_J]=1.$
\item Chain Rule:
\begin{equation*}Pr_\Sigma[E_{i_{1}},..., E_{i_{k}}|\mathbb{E}_L]
=\prod_{l=1}^{k}Pr_\Sigma[E_{i_{l}}|\mathbb{E}_L, E_{i_1},...,
E_{i_{l-1}}].
\end{equation*}
\item Total Probability Rule: If $J,K,L$ are subsequences of $(n]$ and $1\leq i\leq n$ such that $J<i<K<L$, and $\{A_l\}$ is a partition of
$spec(M_i)$, 
then we have:
\begin{equation*}Pr_\Sigma[\mathbb{E}_J,\mathbb{E}_K,\mathbb{E}_L]
=\sum_l Pr_\Sigma[\mathbb{E}_J,\{M_i\in A_l\},\mathbb{E}_K]
\cdot
Pr_\Sigma[\mathbb{E}_L|\mathbb{E}_J,\{M_i\in
A_l\},\mathbb{E}_K].
\end{equation*}
\end{enumerate}
\end{prop}

As shown in Subsec.~\ref{state}, only a partial monotonicity is
valid, and the total probability law does not hold for conditional
probability in a state. Propositions~\ref{con-p1}.2
and~\ref{con-p1}.5 shows that both the full monotonicity and the
total probability law can be recovered for conditional probability
in a test.

\subsection{Independence}

With quantum conditional probability, we are able to introduce the notion of independence for quantum events.

\begin{defn}Given a test $\Sigma=(\rho;M_1,M_2,...,M_n)$. Let $E_i$ be an event defined by $M_i$ for each $1\leq
i\leq n$, let $K$ be a subsequence of $(n]$, let $K<i\leq n$, and let $J$ be a subsequence of $K$. We say that $E_i$ is independent of $\mathbb{E}_J$ with respect to $\mathbb{E}_{K\setminus J}$ in $\Sigma$ if $Pr_\Sigma[E_i|\mathbb{E}_K]=Pr_\Sigma[\mathbb{E}_i|\mathbb{E}_{K\setminus J}].$ In this case, we write $\mathit{Ind}_\Sigma(E_i|\mathbb{E}_J;\mathbb{E}_{K\setminus J})$. 

In particular, if $J=K$, i.e. $\mathit{Ind}_\Sigma(E_i|\mathbb{E}_K;\mathbb{E}_\emptyset)$ and $Pr_\Sigma[E_i|\mathbb{E}_K]=Pr_\Sigma[E_i]$, then we say that $E_i$ is independent of $\mathbb{E}_K$ in $\Sigma$ and simply write $\mathit{Ind}_\Sigma(E_i|\mathbb{E}_K)$. 
\end{defn}

The following are some basic properties of independence relation.

\begin{prop}\label{ind-prop}Given a test $\Sigma=(\rho;M_1,M_2,...,M_n)$. We have:
\begin{enumerate}
\item  $\mathit{Ind}_\Sigma(I_{M_i}|\mathbb{E}_J;\mathbb{E}_{K\setminus J})$ for all $i, K$ and $J$.
\item If $\mathit{Ind}_\Sigma(E_i|\mathbb{E}_J;\mathbb{E}_{K\setminus J})$, then $\mathit{Ind}_\Sigma(\overline{E_i}|\mathbb{E}_J;\mathbb{E}_{K\setminus J})$.
\item If $j$ is an element of $K$, then $\mathit{Ind}_\Sigma(E_i|E_j;\mathbb{E}_{K\setminus j})$ implies $\mathit{Ind}_\Sigma(E_i|\overline{E_j};\mathbb{E}_{K\setminus j})$.
\item If $E_i^{(t)}=\{M_i\in A_t\}$ $(t=1,2)$ with $A_1\cap A_2=\emptyset,$ and $\mathit{Ind}_\Sigma(E_i^{(t)}|\mathbb{E}_J;\mathbb{E}_{K\setminus J})$ $(t=1,2)$, then $\mathit{Ind}_\Sigma(E_i^{(1)}\cup E_i^{(2)}|\mathbb{E}_J;\mathbb{E}_{K\setminus J})$. 
\end{enumerate}\end{prop}

There are some essential difference between independence relations for classical and quantum events. 
First, $\mathit{Ind}_\Sigma(E_{k+1}|I_{M_1},...,I_{M_k})$ does not hold in general. For example, let $\Sigma=(\rho;M_1,M_2)$, where $\rho=|+\rangle\langle +|$, and $M_1,M_2$ are defined by Eq.~(\ref{ex-mea}). Then $\mathit{Ind}_\Sigma(\{M_2=0\}|I_{M_1})$ is not true because $Pr_\Sigma[\{M_2=0\}|I_{M_1}]=\frac{1}{2}\neq 1=Pr_\Sigma[\{M_2=0\}]$. Second, in the classical probability theory, we know that event $E_1$ is independent of $E_2$ if and only if $E_2$ is independent of $E_1$. However, we cannot talk $\mathit{Ind}_\Sigma(E_i|E_j)$ and $\mathit{Ind}_\Sigma(E_j|E_i)$ simultaneously because whenever $\mathit{Ind}_\Sigma(E_i|E_j)$ is well-defined, then $i>j$ and $\mathit{Ind}_\Sigma(E_j|E_i)$ is not well-defined, and vice versa.

\section{QUANTUM LOV\'{A}SZ LOCAL LEMMA}\label{ql}

To give a compact presentation of quantum LLL, we need the following technical definition.

\begin{defn}Given a test $\Sigma=(\rho;M_1,M_2,...,M_n)$ and a sequence $\mathbb{E}=E_1,E_2,...,E_n$ of
events such that $E_i$ is defined by measurement $M_i$ for every
$1\leq i\leq n$.
\begin{enumerate}\item Let $K$ be a subsequence of $(n]$ and $K<i\leq n$. If $\mathit{Ind}_\Sigma(E_i|\overline{\mathbb{E}_K})$, where $\overline{\mathbb{E}_K}$ is the sequence of events obtained by substituting event $E_j$ with $\overline{E_j}$ for each $j$ in $K$, then we say that $E_i$ is negatively independent of $\mathbb{E}_K$ in $\Sigma$ and write $\mathit{NInd}_\Sigma(E_i|\mathbb{E}_K)$.
 \item Let $1\leq l<k\leq n$. We say that $k$ is
negatively independent of $l$ in $(\Sigma,\mathbb{E})$ if
$\mathit{NInd}_\Sigma(E_k|E_1,E_2,...,E_j)\ {\rm for\ all}\ 1\leq j\leq l.$ In
this case, we write $\mathit{NInd}_{\Sigma,\mathbb{E}}(k|l)$. Otherwise, we say that $k$ is
negatively dependent on $l$ in $(\Sigma,\mathbb{E})$ and write $\mathit{NDep}_{\Sigma,\mathbb{E}}(k|l)$.\end{enumerate}
\end{defn}

Now we are ready to present a general version of quantum LLL.

\begin{thm}\label{glll} (General Quantum Lov\'{a}sz Local Lemma)

Let $\mathbb{E}=E_1,...,E_n$ be a sequence of events in Hilbert space
$\mathcal{H}$, where $E_i$ is defined by measurement $M_i$ for every
$i=1,2,...,n$, let $\rho$ be a density operator in $\mathcal{H}$,
and let $x_1,x_2,...,x_n\in (0,1]$. We consider test
$\Sigma=(\rho;M_1,M_2,...,M_n)$. If for all $1\leq i\leq n$,
\begin{equation}\label{assu} Pr_\Sigma[E_i]\leq
x_i\cdot\prod_{j=s_i+1}^{i-1}(1-x_j)
\end{equation} where $s_i=\max\{j<i: \mathit{NInd}_{\Sigma,\mathbb{E}}(i|j)\}.$ Then we have
$Pr_\Sigma[\overline{E_1},...,\overline{E_n}]\geq\prod_{i=1}^{n}(1-x_i).$
\end{thm}

The above theorem can be proved by a procedure similar to that used
both in~\cite{AS04} for the classical LLL and in~\cite{AKS10} for
the relative dimension version of QLLL. We first prove the following
key lemma.

\begin{lem}\label{k-lem}We assume the conditions of Theorem~\ref{glll}. Then for each $1\leq i\leq n$, we have
$Pr_\Sigma[E_i|\overline{E_1},...,\overline{E_{i-1}}]\leq x_i.$
\end{lem}

\textit{Proof.} We proceed by induction on $i$. For the base case of
$i=1$, it follows immediately from the assumption~(\ref{assu}) that
$Pr_\Sigma[E_i|\overline{E_1},...,\overline{E_{i-1}}]=Pr_\Sigma[E_1]\leq
x_1.$ Now we assume that the conclusion holds for all the cases up
to $i-1$. If $s_i=i-1$, then $\mathit{NInd}_{\Sigma,\mathbb{E}}(i|i-1)$ and
$\mathit{NInd}_\Sigma(E_i|\overline{E_1},...,\overline{E_{i-1}})$. This
together with the assumption~(\ref{assu}) leads to
$Pr_\Sigma[E_i|\overline{E_1},...,\overline{E_{i-1}}]=Pr_\Sigma[E_i]\leq
x_i.$ For $k<i-1$, we have:
$$Pr_\Sigma[E_i|\overline{E_1},...,\overline{E_{i-1}}]=\frac{Pr_\Sigma[\overline{E_{s_i+1}},
...,\overline{E_{i-1}},E_i|\overline{E_1},\overline{E_2},...,\overline{E_{s_i}}]}{Pr_\Sigma
[\overline{E_{s_i+1}},
...,\overline{E_{i-1}}|\overline{E_1},\overline{E_2},...,\overline{E_{s_i}}]}.$$
Then we need to examine the numerator and denominator of the
fraction on the right hand side of the above equation. For the
numerate, it holds that
\begin{equation}\label{nu}\begin{split}
Pr_\Sigma[\overline{E_{s_i+1}},
...,\overline{E_{i-1}},E_i&|\overline{E_1},...,\overline{E_{s_i}}]\leq
Pr_\Sigma[E_i|\overline{E_1},...,\overline{E_{s_i}}]\\
& =Pr_\Sigma[E_i]\leq x_i\cdot\prod_{j=s_i+1}^{i-1}(1-x_j)
\end{split}\end{equation} because the definition of $s_i$ implies
$\mathit{NInd}_\Sigma(E_i|E_1,...,E_{s_i})$. Here, the first inequality follows
from Proposition~\ref{con-p}.2, and the last inequality is exactly
the assumption~(\ref{assu}). For the denominator, we first notice
that the induction hypothesis yields
$Pr_\Sigma[E_j|\overline{E_1},...,\overline{E_{s_i}},\overline{E_{s_i+1}},...,\overline{E_{j-1}}
]\leq x_j$ for all $j\leq i-1$. Therefore, by the Chain Rule
(Proposition~\ref{con-p}.4) we obtain:
\begin{equation}\label{de}\begin{split}
Pr_\Sigma[\overline{E_{s_i+1}},...,\overline{E_{i-1}}|\overline{E_1},...,\overline{E_{s_i}}]
&=\prod_{j=s_i+1}^{i-1}Pr_\Sigma[\overline{E_j}|\overline{E_1},...,\overline{E_{s_i}},\overline{E_{s_i+1}},...,
\overline{E_{j-1}}]\\ &\geq \prod_{j=s_i+1}^{i-1}(1-x_j).
\end{split}\end{equation}Finally, we combine Eqs.~(\ref{nu})
and~(\ref{de}) and assert that the conclusion holds in the case of
$i$. This completes the proof. $\Box$

\textit{Proof of Theorem~\ref{glll}}. It follows from
Lemma~\ref{k-lem} and Proposition~\ref{con-p}.3 that
$Pr_\Sigma[\overline{E_i}|\overline{E_1},...,\overline{E_{i-}}]\geq
1-x_i$ for all $1\leq i\leq n$. Then by the Chain Rule we obtain:
\begin{equation*}
Pr_\Sigma[\overline{E_1},...,\overline{E_n}]=\prod_{i=1}^{n}Pr_\Sigma
[\overline{E_i}|\overline{E_1}, ...,\overline{E_{i-1}}] \geq
\prod_{i=1}^{n}(1-x_i).\ \Box\end{equation*}

A symmetric version of QLLL immediately follows from the general QLLL. To present it in a concise way, we first introduce the following technical definition.

\begin{defn}Let $\Sigma=(\rho;M_1,M_2,...,M_n)$ be a test, and let $\mathbb{E}=E_1,E_2,...,E_n$ be a sequence of events such that
$E_i$ is defined by measurement $M_i$ for each $1\leq i\leq n$. A
nonnegative integer $d$ is called a dependence radius of sequence
$E_1,E_2,...,E_n$ in $\Sigma$ if for any $1\leq l<k\leq n$,
$\mathit{NDep}_{\Sigma,\mathbb{E}}(i|j)\ {\rm implies}\ l\geq k-d.$\end{defn}

\begin{thm}\label{slll} (Symmetric Quantum Lov\'{a}sz Local Lemma)

Given a test $\Sigma=(\rho;M_1,M_2,...,M_n)$. Let $E_1,...,E_n$ be a
sequence of events such that $E_i$ is defined by measurement $M_i$
for every $1\leq i\leq n$, and let $d$ be a dependence radius of
sequence $E_1,E_2,...,E_n$ in $\Sigma$. If $Pr_\Sigma[E_i]\leq p$
for all $1\leq i\leq n$, and $p\cdot e\cdot (d+1)\leq 1$, then
$Pr_\Sigma[\overline{E_1},...,\overline{E_n}]>0$.
\end{thm}

\textit{Proof.} We first observe that $s_i\geq i-d-1$ for all $1\leq
i\leq n$ whenever $d$ is a dependence radius of the sequence
$E_1,...,E_n$. Now put $x_i=\frac{1}{d+1}$ for all $1\leq i\leq n$.
Then we have
\begin{equation*}\begin{split}
Pr_\Sigma[E_i]\leq p\leq \frac{1}{(d+1)\cdot e}\leq
\frac{1}{d+1}(1-\frac{1}{d+1})^{d}\leq
x_i\cdot\prod_{j=s_i+1}^{i-1}(1-x_j).
\end{split}\end{equation*}Thus, it follows from Theorem~\ref{glll}
that
$Pr_\Sigma[\overline{E_1},...,\overline{E_n}]\geq\prod_{i=1}^{n}(1-x_i)>0.\
\Box$

It is worth noting that in classical probability, the ordering of
events in a sequence can be freely changed. So, in the
assumption~(\ref{assu}), for each $i$, the sequence $E_1,...,E_i$
can be arranged in such a way: those events independent of $E_i$ are
far from $E_i$, and the events that $E_i$ depends on are close to
$E_i$. Thus, the general LLL (see~\cite{EL75}; \cite{AKS10}, Theorem
12) and the symmetric LLL can be recovered in some cases from
Theorems~\ref{glll} and~\ref{slll}, respectively.

\section{CONCLUSION}

This paper presents the quantum generalization of both the general and the symmetric Lov\'{a}sz Local Lemma (LLL) for a sequence
of measurements. In fact, these results even hold in the dynamic case where unitary evolutions or even general quantum operations
modeled by trace-preserved completely positive operators
are allowed between these measurements.

The measurements in a test considered in this paper are totally ordered in a sequence. An interesting open problem for further studies is how to generalize the results obtained here to the case where the measurements in a test are only partially ordered. 

\textbf{Acknowledgement:} The author is very grateful to Professors Stan Gudder and Yuan Feng and Dr. Or Sattath for pointing out a serious mistake in the original version of this paper. He also wishes to thank Professor Masanao Ozawa for a very helpful discussion.

\end{document}